\documentclass[conference]{IEEEtran}
\IEEEoverridecommandlockouts
\usepackage{cite}
\usepackage{amsmath,amssymb,amsfonts}
\usepackage{algorithmic}
\usepackage{graphicx}
\usepackage{textcomp}
\usepackage{xcolor}
\usepackage{url}
\usepackage[shortlabels]{enumitem}
\usepackage{caption}
\usepackage{subfigure}
\usepackage{setspace}
\usepackage[hidelinks]{hyperref}
\def\BibTeX{{\rm B\kern-.05em{\sc i\kern-.025em b}\kern-.08em
    T\kern-.1667em\lower.7ex\hbox{E}\kern-.125emX}}

\begin{document}

\title{Visual Analysis of GitHub Issues to Gain Insights%
}

\author{\IEEEauthorblockN{Rifat Ara Proma}
\IEEEauthorblockA{\textit{Scientific Computing and Imaging Institute} \\
\textit{University of Utah}\\
Salt Lake City, UT, USA \\
proma@sci.utah.edu}
\and
\IEEEauthorblockN{Paul Rosen}
\IEEEauthorblockA{\textit{Scientific Computing and Imaging Institute} \\
\textit{University of Utah}\\
Salt Lake City, UT, USA \\
prosen@sci.utah.edu}
}

\maketitle

\begin{abstract}
Version control systems are integral to software development, with GitHub emerging as a popular online platform due to its comprehensive project management tools, including issue tracking and pull requests. However, GitHub lacks a direct link between issues and commits, making it difficult for developers to understand how specific issues are resolved. Although GitHub's Insights page provides some visualization for repository data, the representation of issues and commits related data in a textual format hampers quick evaluation of issue management. This paper presents a prototype web application that generates visualizations to offer insights into issue timelines and reveals different factors related to issues. It focuses on the lifecycle of issues and depicts vital information to enhance users' understanding of development patterns in their projects. We demonstrate the effectiveness of our approach through case studies involving three open-source GitHub repositories. Furthermore, we conducted a user evaluation to validate the efficacy of our prototype in conveying crucial repository information more efficiently and rapidly.
Video URL : \texttt{\small\href{https://youtu.be/bFrMGfwax68}{https://youtu.be/bFrMGfwax68}}
\end{abstract}


\section{Introduction}

\begin{figure*}[!t]
    \centering
    {\includegraphics[trim=78pt 60pt 53pt 25pt, clip, width=0.975\linewidth]{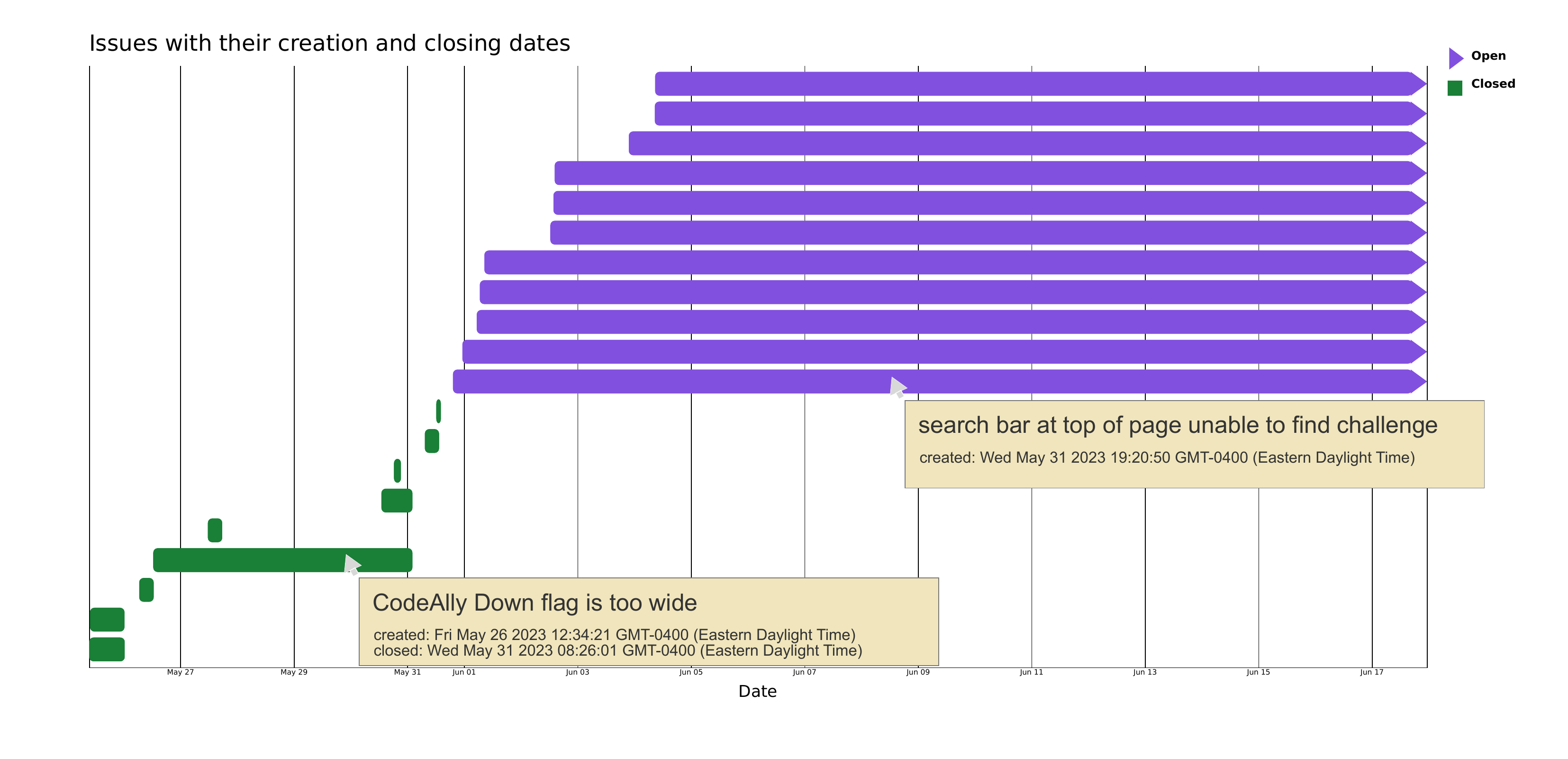}}
    
    \caption{\textsc{Timeline View} illustrating the status of recent issues from the \textit{freeCodeCamp} repository. The horizontal axis represents the date, while the bars' width corresponds to the issues' duration. Open issues are colored purple, while closed issues are colored green. Furthermore, open issues are accompanied by an arrow on the right to signify their ongoing nature and continuity. Additional information is revealed through tooltips when hovering the mouse over the bars.}
    \label{timeline-freecodecamp}
\end{figure*}

Collaboration between developers is one of the most critical tasks in software development. Distributed version control systems, such as Git, allow programmers to store source code, collaborate, and track their work. GitHub is one of the most widely used online Git-based platforms, with over 100 million developers and more than 420 million repositories as of February 2024~\cite{GitHubWiki}. GitHub offers several useful functionalities, e.g., issue tracking, pull requests, etc., that make the collaborative development process more accessible. Issue tracking, in particular, helps keep track of work done, what work needs to be done, and who is working on what. It is used for several purposes, including reporting bugs, feature requests, and enhancements.

There are many existing work in the field of GitHub data mining with many different applications, including predicting topics for repositories~\cite{zhou2021ghtrec}, understanding user behaviors~\cite{chatziasimidis2015data}, predicting defects in softwares~\cite{suhag2022software}. Many of the existing works have also proposed using data visualization to enhance different aspects of GitHub~\cite{dubey2018data, izquierdo2015gila, celinska2017programming}. In this paper, we have identified a problem with the current GitHub interface and attempted to solve it using data visualization.

Despite GitHub's numerous functionalities, including an insights page that utilizes visualizations to provide insights about the repository, its presentation of information regarding issues, commits, and other activities is primarily textual. This format can limit the ability to discern patterns and extract deeper insights. Data visualization effectively conveys the meaning of additional data through charts. Visualizing GitHub issues and commits makes it possible to extract meaningful information from them and use it to analyze patterns and problems in software projects. For instance, visual representations can highlight trends in issue resolution times, reveal which components of the software are most prone to bugs, and track the contributions of individual developers more clearly.

This paper presents a prototype web application that shows GitHub issue data through several visualizations. Each one offers different types of insights into the repository issues and related commits, accomplishing tasks that include:
\begin{itemize}
\item Facilitating programmers in understanding the duration of an issue, both in terms of how long it has been open and the time it took to resolve;
\item Gaining insights regarding different types of issues based on labels assigned;
\item Narrowing which commits are responsible for resolving an issue;
\item Checking which files were updated to resolve issues; and
\item Determining which files are getting more updates to resolve issues.
\end{itemize}

Our approach aimed to enable developers to identify bottlenecks in the development process quickly, recognize high-risk areas of the codebase and understand the dynamics of team contributions. In turn, supporting more strategic planning and more effective issue management.

We collected data about three public repositories and analyzed them using our application to evaluate our approach. Additionally, we conducted a small-scale expert user study involving four participants that validated the efficiency of the proposed solution. The study involved developers using our visualizations to answer some questions about the repositories, and the feedback we received indicated that our tool significantly enhanced their ability to interpret and act upon the data from GitHub. This validation emphasizes the practical benefits of integrating data visualization into version control systems, making it a promising direction for further research and development.

\section{Related Work}
Considering the expanding popularity of GitHub and its open API, several existing research studies have delved into analyzing different aspects of it to enhance different purposes. This section discusses prior work regarding analyzing GitHub projects and using data visualizations to enhance some aspects.

\subsection{Analysis of GitHub Projects}
A lot of existing research collects data about GitHub repositories and uses it to enhance user experience. Zhou et al.~\cite{zhou2021ghtrec} presented a GHTRec service that predicts topics for GitHub repositories and then gives users personalized recommendations for trending repositories. Analyzing GitHub data can also help to understand user behavior and communication patterns among collaborators~\cite{chatziasimidis2015data, ortu2018mining}. It can also enable detecting defective data~\cite{suhag2022software} and factors that impact project success~\cite{chatziasimidis2015data}. Such approaches can be useful for software development and testing teams to shorten the development life cycle. 
 
Analyzing GitHub data can potentially improve programming languages like Java. Lemay~\cite{lemay2019understanding} inspected publicly available Java source code that could infer usability issues with the Java language. Zhang et al.~\cite{zhang2018multiple} also studied 9,476 competing pull requests from 60 Java repositories on GitHub to analyze how multiple pull requests update the same code. 

Researchers have also explored machine learning-based approaches to enhance GitHub. Montandon et al.~\cite{montandon2021mining} proposed a method to automatically recognize the technical role of open source developers, which can help to form cross-functional teams. Ticket Tagger~\cite{kallis2021predicting} also used machine learning techniques to evaluate issue titles and descriptions and automatically assign labels to each issue. However, none of these existing works explored utilizing data visualization to enhance different aspects of GitHub.

\subsection{Visualization of GitHub Projects}
Some existing research has also explored visualization of data retrieved from GitHub to achieve different objectives. Dubey et al.~\cite{dubey2018data} visualized repository parameters collected from GitHub to identify the most popular domains and languages on GitHub. Celinska et al.~\cite{celinska2017programming} proposed a visualization with a weighted network of programming languages used on GitHub to help developers choose programming languages for their applications. Visualizations have also been utilized in tools like GiLA~\cite{izquierdo2015gila} to categorize issues based on labels and analyze developer locations and the technology they use~\cite{rusk2014location}.

A web-based tool, ChangeViz, was proposed by Gasparini et al.~\cite{gasparini2021changeviz} to help developers better understand GitHub pull requests. The tool combines the existing interface of GitHub with the visualization approach of STACKSPLORER~\cite{karrer2011stacksplorer}. Another visualization tool proposed by Højelse et al.~\cite{hojelse2022git} offered multiple views for analyzing the evolution of hierarchically organized Git repositories. MyGitIssues~\cite{ristemi2019mygitissues} is a web application that lists issues of specific GitHub repositories using data visualization approaches. It makes managing different repositories easier with the support of technologies such as WebSockets and WebHooks. Fiechter et al.~\cite{fiechter2021visualizing} also worked on presenting issues through visualization by introducing an approach called issue tales, which is an interactive visual analytics tool. It allows for analyzing all events and actors related to any GitHub issue, such as related commits, pull requests, related issues, etc. However, it does not provide any way to get an overall overview of different issue types and their timelines. The issue-related behavior of GitHub was also investigated via visualization techniques by Liao et al.~\cite{liao2018exploring}. Their analysis emphasized the importance of issue labeling and issue-related behavior analysis for better issue-handling practices. They mentioned that the time distribution of issue commits maintains a development model that roughly corresponds to the project's life cycle.

Numerous studies have been conducted on GitHub data mining and visualization, particularly in understanding GitHub issues. However, existing approaches lack an interface that provides an overview of multiple issues at a glance and offers interactive capabilities for identifying related commits and file changes. This restricts the identification of trends within a repository. We intended to fill this gap by developing interactive visualization interfaces that empower developers to gain deeper insights into issue trends and explore their repositories more effectively. This paper proposes methods for visualizing GitHub issues to enhance developers' understanding of issues and commit management within their repositories.

\section{Our Approach}

To address the gap in the available tools for evaluating GitHub issues, we developed a web-based analysis tool, using D3.js~\cite{bostock2011d3}, to generate visualizations. The code for the prototype web application can be accessed at \texttt{\href{https://github.com/RifatAraProma/GitHubDataViz}{GitHubDataViz}}\footnote{\url{https://github.com/RifatAraProma/GitHubDataViz}}. The web application can also be accessed at \texttt{\href{https://github-issue-analyzer-d967eeb35538.herokuapp.com/}{GitHub Issue Analyzer}}\footnote{\url{https://github-issue-analyzer-d967eeb35538.herokuapp.com/}}.

\subsection{Design Considerations}
\label{sec:design}
This work aimed to create a useful resource that enables developers to evaluate their repositories better and comprehend development trends. Initially, we identified that the current GitHub issue does not allow the exploration of issues in a visual format and proposed a series of visualizations that can enhance the process. To ensure the proposed visualization tool enables developers to analyze issues and related commits in a repository, we interviewed two software engineers to understand their requirements. The interview was open-ended, and we asked them how they utilize the current GitHub interface to manage their repositories. Specifically, we asked them how they identify issues that have been open for a long time, identify what resolved a particular issue, which files were changed during the process, and what will help them perform these tasks efficiently. After that, we showed them some of our designed visualization solutions and asked for their feedback. These interviews played a crucial role in shaping the design of our visualization solution. The software engineers actively participated throughout the design process, providing valuable insights and feedback. By incorporating their expertise and feedback, we ensured that the resulting visualization tool aligned with their needs and preferences. This collaborative approach allowed us to create a tool that facilitates the extraction of meaningful insights regarding issues and their associated commits within a repository.

We aimed to design a collection of interactive interfaces that enable the visualization of information about issues, commits, and updated files, addressing the following tasks:

\begin{enumerate}[start=0,label={T\arabic*:}]
    \item Displaying the timeline of issues in a repository, showcasing their status and duration in each state.
    \item Providing a clearer understanding of the types of issues present in a repository.
    \item Visualizing various aspects of an issue concisely and empowering developers to identify the commits associated with a specific issue and the files that were modified during the resolution process.
    \item Offering visualizations that facilitate comprehension of file update frequencies within a repository, specifically focusing on bug fixes. 
\end{enumerate}

Throughout the remainder of the paper, we have specified when a software feature or evaluation supports one of these tasks by denoting the task identifier inside parentheses. By addressing these objectives, our visualization tool seeks to enhance the effectiveness and efficiency of repository analysis.

\begin{figure*}[!tbp]
    \centering

    \subfigure[\textsc{Timeline View} showing labels of issues\label{label-freecodecamp}]{\includegraphics[trim=40pt 25pt 10pt 50pt, clip, height=100pt]{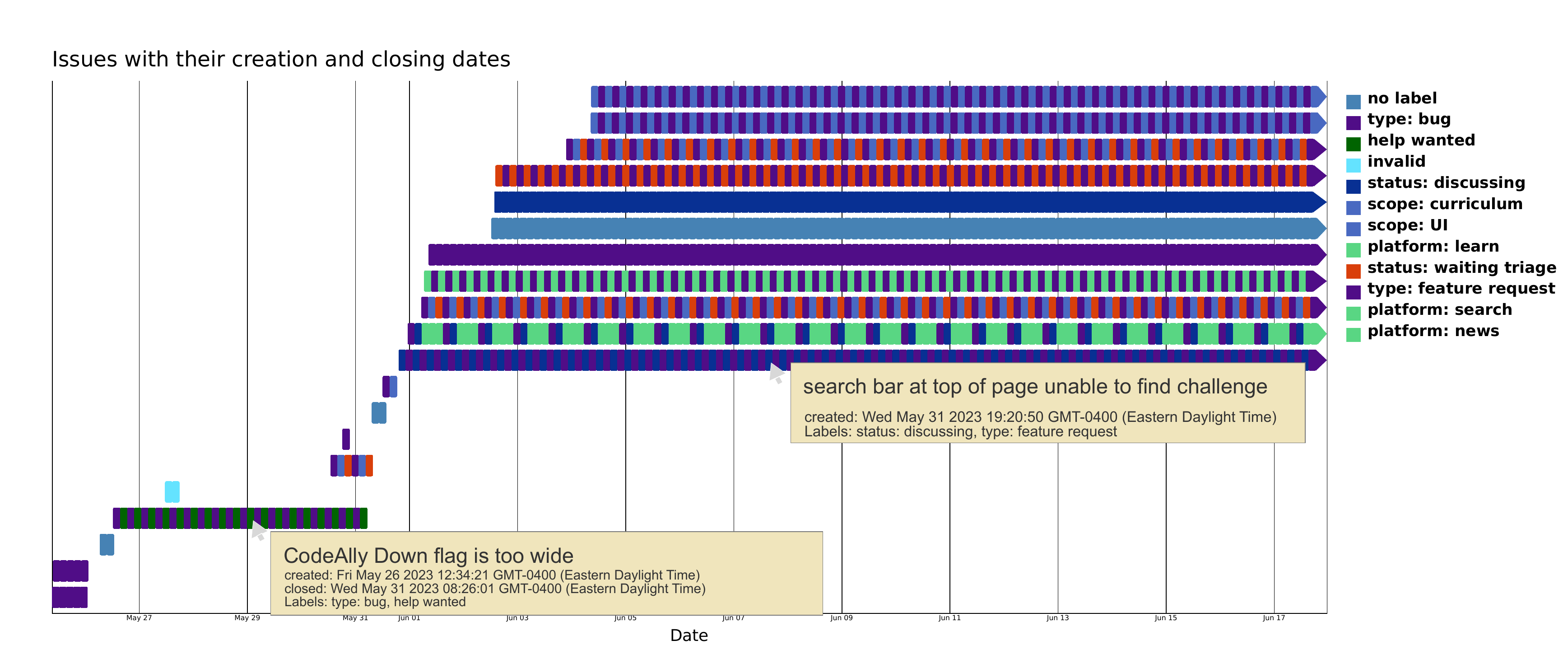}}
    \hspace{5pt}
    \subfigure[\textsc{Issue Graph} of the longest resolved issue\label{graph-freecodecamp-simple}]{\includegraphics[trim=75pt 10pt 75pt 115pt, clip, height=95pt]{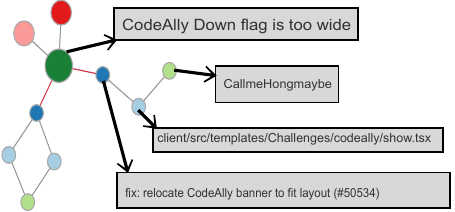}\includegraphics[trim=265pt 140pt 40pt 15pt, clip, height=50pt]{Figure/freeCodeCamp/freeCodecamp_network_closed_issue_annotated.pdf}}

    \subfigure[\textsc{Summary of Updated Files} of the repository\label{donut-freecodecamp}]{
        \includegraphics[height=110pt]{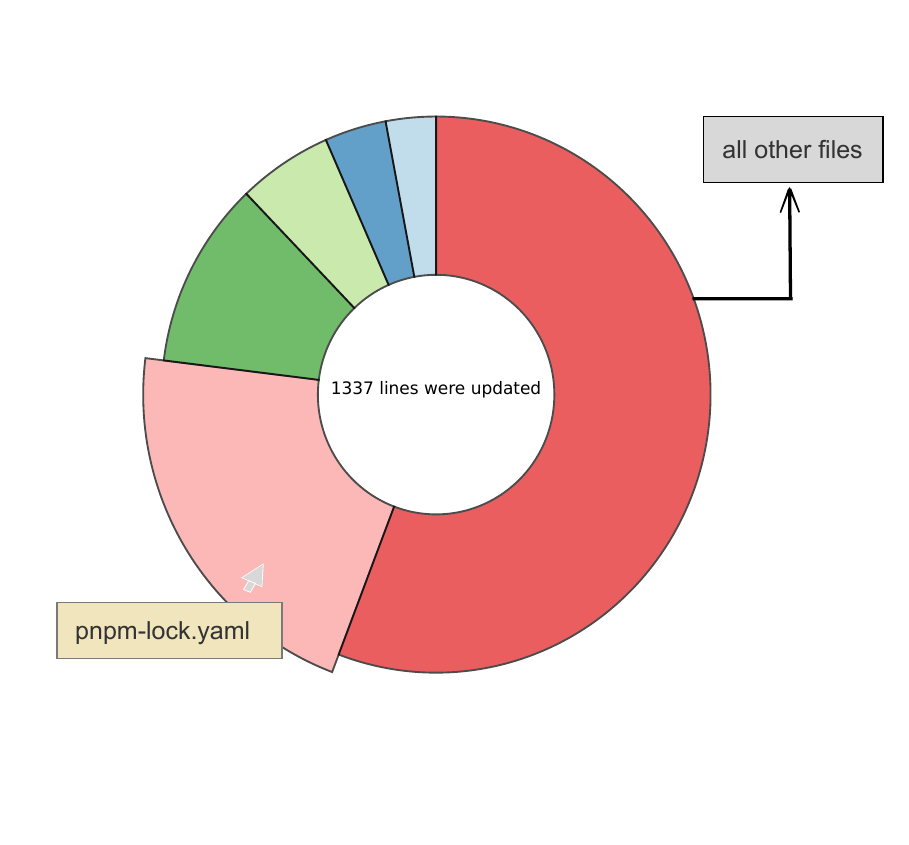}
        \hspace{5pt}
        \includegraphics[width= 0.5\linewidth]{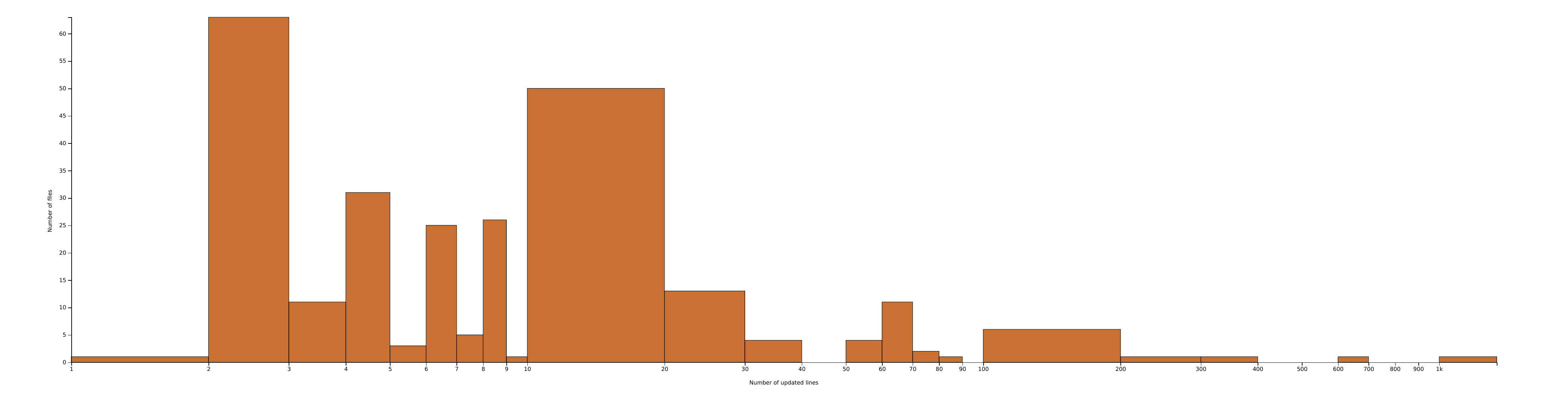}
        \hspace{5pt}
        \includegraphics[height=110pt]{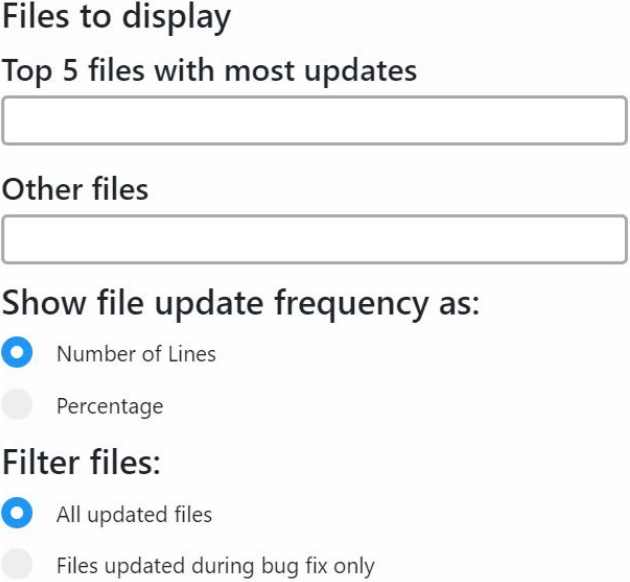}
     }

    \caption{Visualization of the issues created between May 24, 2023 - June 18, 2023, in \textit{freeCodeCamp} repository using our approach. The \textsc{Timeline View}, depicted in (a), showcases the associated label colors of the issues through alternating striped bars. (b) displays the \textsc{Issue Graph} corresponding to a closed issue with the longest resolution time. (c) shows the \textsc{Summary of Updated Files} view. On the left is a donut chart illustrating the top 5 files with the highest number of updated lines, while the sixth wedge represents the cumulative sum of updated lines in the remaining files. In the middle, a histogram provides insights into the distribution of files across different ranges of updates. Finally, the interface on the right enables users to filter or modify the donut chart based on their preferences. Gray boxes represent annotations in all the figures, while a mouse icon and light yellow boxes indicate tooltips. }

\end{figure*}

\subsection{Data Collection}
We developed a custom Python script for this project to extract relevant data from a GitHub repository. The script retrieved information about recent activities, including issue details such as title, assignees, labels, opening and closing times, creator and closer users, commit messages, commit timestamps, updated files, and commit authors. Specifically, the script focused on designated repositories.

We employed the GitHub API to fetch information about the 100 most recent issues. Subsequently, we filtered out any issues identified as pull requests, as they were irrelevant to our analysis. In addition, we utilized the GitHub API to access the Commits page on the GitHub user interface, allowing us to retrieve information about all commits present in the Newer tab of the page. The Python script leveraged the API to retrieve and store the desired data in a CSV file. This data serves as the foundation for generating visualizations at a later stage.

\subsection{Interfaces}
The tool we constructed consists of 3 main views. Initially, it presents users with the option to select a repository. Upon selection, a \textsc{Timeline View} is displayed (see \autoref{timeline-freecodecamp}), showcasing recent issues along with their corresponding status (open or closed). Users can use a drop-down button to switch the view and display labels associated with the issues to enhance flexibility. By clicking on a bar within the \textsc{Timeline View}, users can investigate the associated issue within the \textsc{Issue Graph} (see \autoref{graph-freecodecamp-simple}). Finally, a button located in the top left corner allows users to switch to the \textsc{Summary of Updated Files} view (see \autoref{donut-freecodecamp}). In the subsequent sections, we provide a comprehensive discussion of each of these interfaces, elucidating their functionalities and features.

\subsubsection{\textsc{Timeline View} (see \autoref{timeline-freecodecamp} and \autoref{label-freecodecamp})}
To visualize the duration of each GitHub issue, a Gantt chart was created due to their effectiveness in showing task schedule~\cite{luz2011comparing}~(T0), as depicted in \autoref{timeline-freecodecamp} and \autoref{label-freecodecamp}. Each issue is represented by a bar, where the horizontal axis corresponded to the date, and the width of the bar represents the number of days the issue remained open. Open issues are displayed in purple color, while closed issues are indicated by green, consistent with GitHub's default color scheme. Open issues extend to the rightmost section of the chart, indicating their ongoing status, and are represented by an arrow to signify continuity. Hovering the mouse over an issue triggers a tooltip that displays the issue title, creation date, closing date, and labels associated with it, providing additional contextual information.

\begin{figure*}[!tbp]
    \centering
    
    \subfigure[\textsc{Timeline View} of issue status\label{timeline-hyprland}]{\includegraphics[trim=75pt 5pt 60pt 0, clip, width=0.475\linewidth]{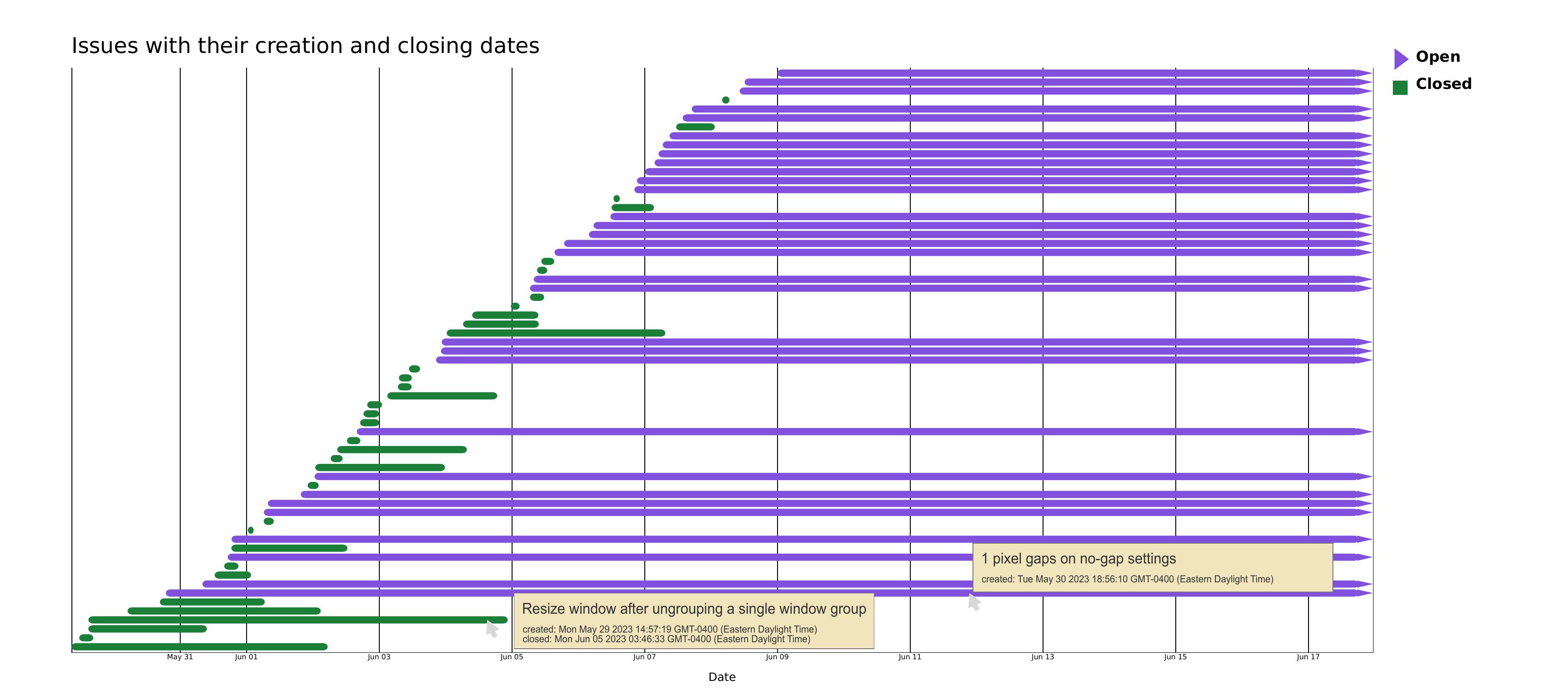}}
    \hfill
    \subfigure[\textsc{Timeline View} of issue labels\label{label-hyperland}]{\includegraphics[trim=75pt 5pt 60pt 0, clip, width=0.475\linewidth]{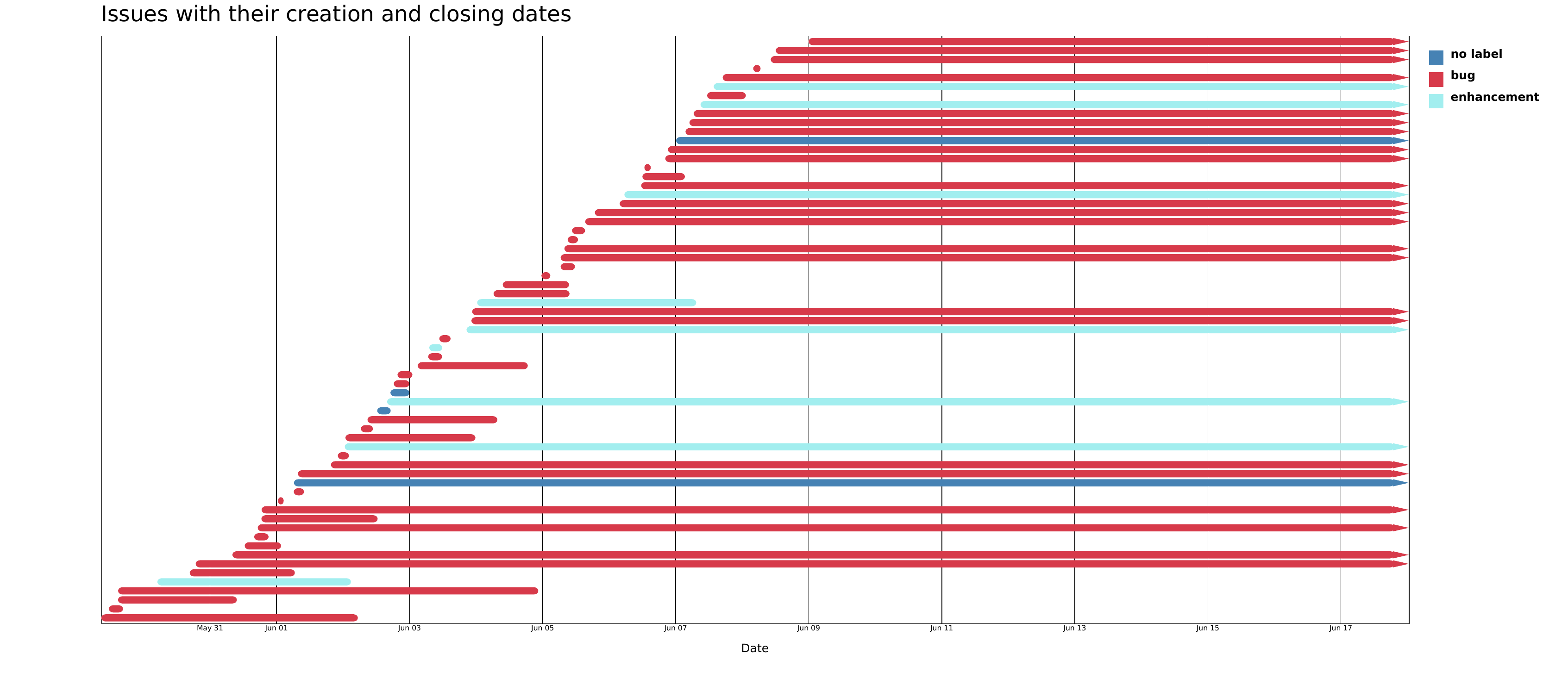}}
    
    \hfill
    \begin{minipage}[t]{0.285\linewidth}
        \centering
        \subfigure[Top five most updated files\label{donut-hyprland}]{\includegraphics[width=.975\linewidth]{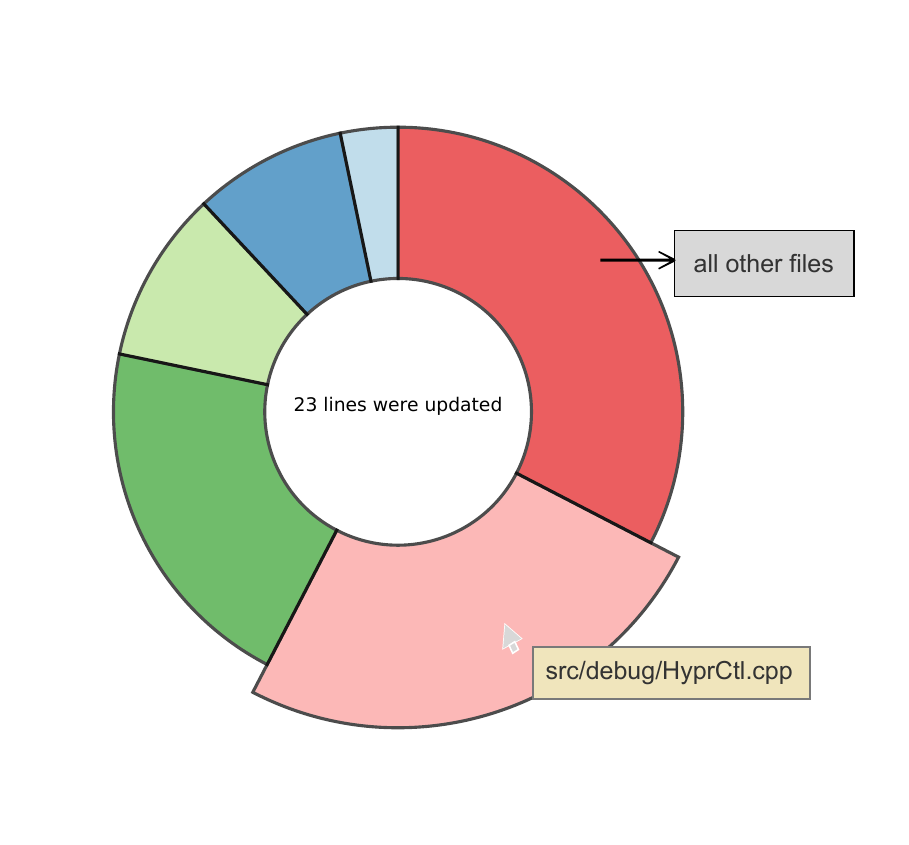}}

        \subfigure[Most updated files for bug fixes\label{donut-bug-hyprland}]{\hspace{5pt}\includegraphics[width=0.975\linewidth]{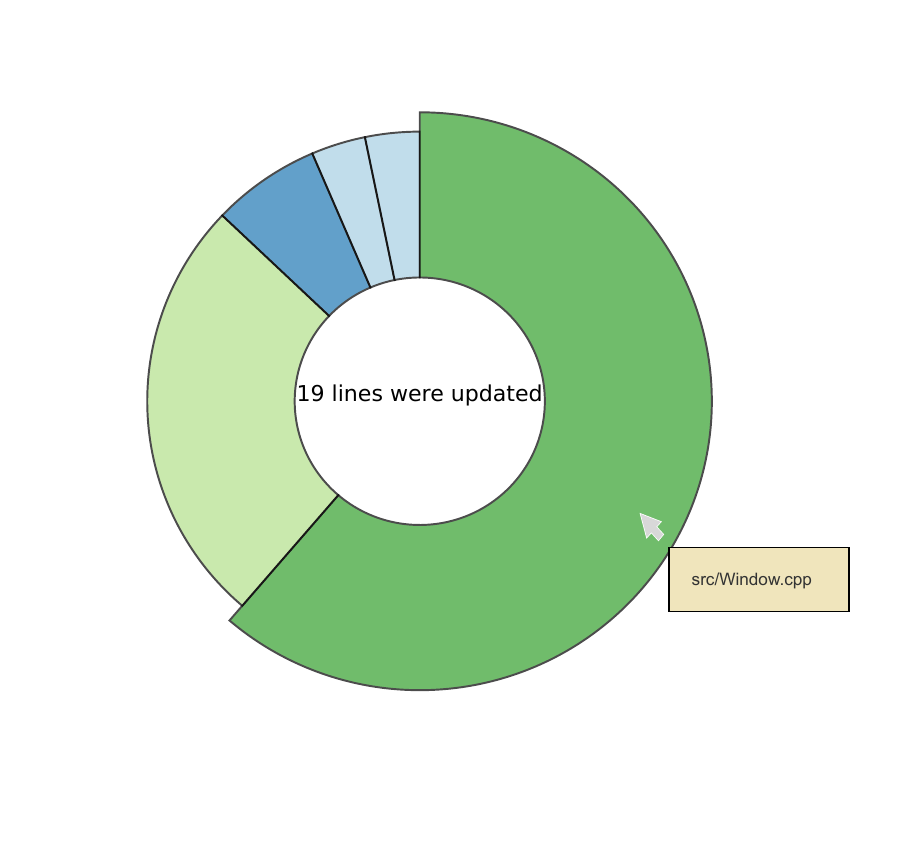}\hspace{5pt}}
    \end{minipage}
    \begin{minipage}[t]{0.705\linewidth}
    \subfigure[\textsc{Issue Graph} of the longest resolved issue\label{graph-hyprland}]{\includegraphics[width=\linewidth]{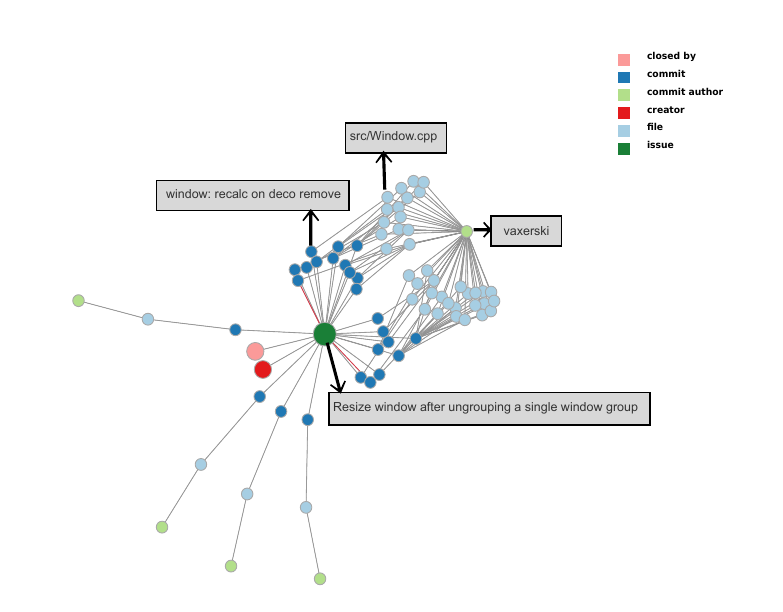}}
    \end{minipage}
    
    \caption{
    Visualizations of the issues created between May 15, 2023, and June 18, 2023, in \textit{Hyprland} repository. The \textsc{Timeline View} shows a significant number of recent issues, where (a) indicates that the majority of the issues remain open, while (b) highlights that a significant portion of them are classified as bug reports. The top updated files are shown in (c) and (d). Finally, the \textsc{Issue Graph}, in (e), provides insight into a closed issue with the longest resolution time. Analyzing the commit messages made it possible to identify the specific commit responsible for fixing the issue, the contributor who made the commit, and the associated updated file.}
    \label{hyprland-all}
\end{figure*}

The \textsc{Timeline View} also incorporates a feature allowing the color-coded display of issue labels~(T1), as shown in \autoref{label-freecodecamp}. One can switch to this view using a drop-down menu on the top left. The colors and names of these labels are specific to the repository being analyzed. The color encoding in this view varies across repositories as shown in \autoref{label-freecodecamp}, \autoref{label-hyperland}, and \autoref{label-javascript}, as the exact colors associated with the labels in each repository are utilized. In case an issue does not have any label, ``steelblue'' color is assigned to it, and the legends mark the color as ``no label". It is possible that a certain repository uses the same color for multiple labels, as shown in \autoref{label-freecodecamp}. Both \texttt{type: bug} and \texttt{type: feature request} use the same purple color. To distinguish two issues having different labels but the same color associated with them, we added labels in the tooltip. In cases where an issue has multiple labels, alternating striped patterns were used to represent the colors of each label.

\subsubsection{\textsc{Issue Graph} (see \autoref{graph-freecodecamp-simple})}
Upon clicking the bars of the \textsc{Timeline View}, a force-directed graph is generated, displaying properties associated with the issue~(T2), as depicted in \autoref{graph-freecodecamp-simple}. The reason behind choosing force-directed graphs for this purpose was their aesthetic~\cite{huang2016effects} and ability to portray the connection between issues and related factors in a natural manner. However, GitHub does not automatically establish connections between commits or pull requests and issues unless they were manually linked or specific keywords were utilized in the pull request to close a particular issue~\cite{GitHubIssueDoc}. Developers often do not utilize these keywords, or manually link pull requests to issues. Consequently, to identify the commits that contributed to the resolution of a specific issue, the search scope is narrowed down to include only the commits made between the opening and closing dates of the issue. Additionally, information such as the issue creator, assignee, and closer is provided in the graph. For each associated commit, nodes are added to represent the files modified in that commit, along with a node representing the commit creator. The color of the issue node remains consistent with the issue colors employed in the \textsc{Timeline View}, reflecting the issue's status. Other colors employed in the graph were selected using ColorBrewer~\cite{harrower2003colorbrewer}, with the intention of being able to distinguish different types of nodes. \autoref{graph-freecodecamp-complex} displays the \textsc{Issue Graph} representing an open issue, as denoted by the purple-colored node. We included the ``closed by" node in the graph for illustrative purposes, although it does not apply to an open issue. Similarly, the absence of an assignee for this particular issue prompted us to add the corresponding node, demonstrating how the graph would appear if assignees were present. It is important to note that aside from these two nodes, all other chart elements accurately reflect real data.

Upon hovering on each node, a tooltip reveals its details, such as the name of the users, files, commit messages, and issue title. Notably, certain edges linking the issue to the commits are assigned a red color, in line with GitHub's color scheme for bugs. These specific commits are deemed bug fixes. It is customary to use the keyword ``fix" to denote a commit that addresses a bug. Therefore, commits are considered related to bug fixes by checking if commit messages include the keyword. The decision to employ a force-directed graph was motivated by the desire to facilitate the formation of natural clusters during issue evaluation. This graphical representation also allowed users to interact with the graph by dragging nodes, enabling them to explore different parts of the graph at their discretion.

\subsubsection{\textsc{Summary of Updated Files} (see \autoref{donut-freecodecamp})}
From the issue \textsc{Timeline View}, one can switch to a view showing \textsc{Summary of Updated Files} that contains a donut chart that shows how many lines of the files of the repository were updated in the process of resolving issues~(T3), as shown in \autoref{donut-freecodecamp}(left). We chose donut charts as they enable viewers to identify proportions quickly~\cite{quadri2021survey}. The chart shows the top 5 files with the most updated lines in the donut chart by default. One portion of the donut chart shows the total number of updated lines of all the other files except those displayed explicitly. This helps to understand the extent of the updates of a particular file. There are filtering options as shown in \autoref{donut-freecodecamp}(right), using which one can add or remove specific files and analyze how that changes the donut chart. There is also an option to see files that are updated during bug fixes only, enabling users to identify files that are more updated to fix bugs. Upon hovering on each wedge, a tooltip appears with the name of the file and the center of the donut shows how many lines are updated in that file. The colors used for this chart were chosen using ColorBrewer~\cite{harrower2003colorbrewer}.

\begin{figure*}[htbp]
    \centering
    {\includegraphics[trim=30pt 25pt 10pt 105pt, clip, width=0.975\linewidth]{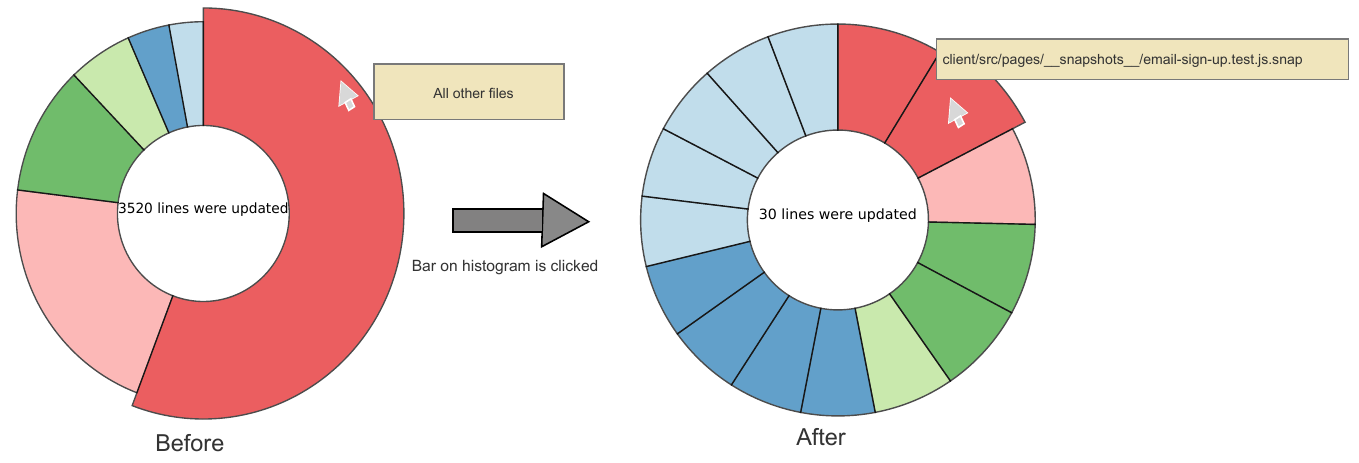}}
    {\includegraphics[trim=20pt 15pt 10pt 25pt, clip, width=0.975\linewidth]{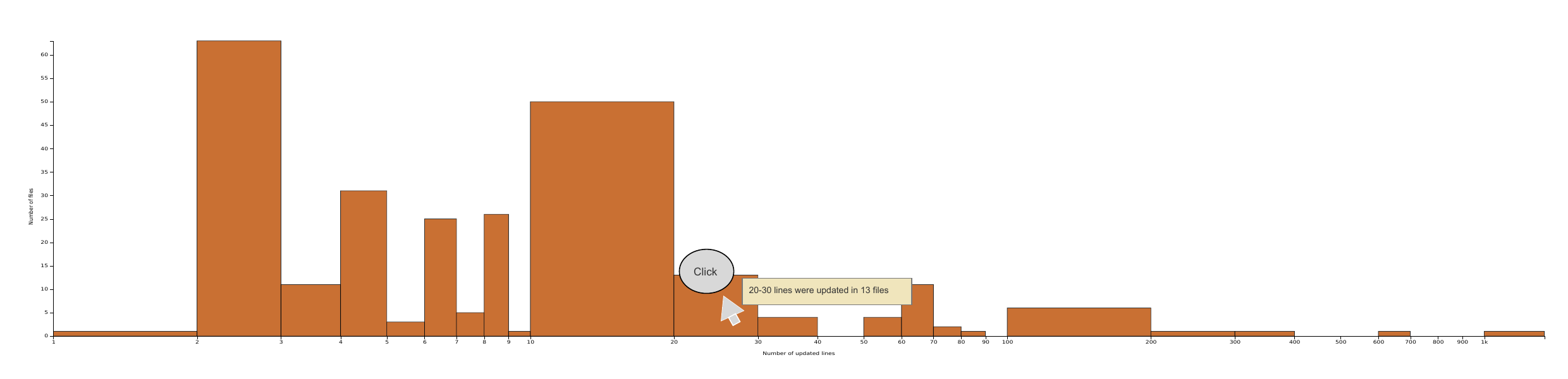}}
    
    \caption{\textsc{Summary of Updated Files} of \textit{freeCodeCamp} repository. The default view shows a donut chart (top left) with the most updated files. After clicking the bar of the histogram (bottom), the donut chart gets updated (top right).}
    \label{donut-histogram-interaction}
\end{figure*}

A histogram that shows the number of files updated, based upon the number of lines modified, as shown in \autoref{donut-freecodecamp}(middle). The horizontal axis of the histogram represents the number of files, and the vertical axis represents the update frequency range on a logarithmic scale. The histogram also lets users interact with the donut chart. Depending on which bar of the histogram is clicked, the donut chart is updated to show files from that update frequency range, as shown in \autoref{donut-histogram-interaction}. This functionality is added to give users the flexibility to explore files of specific update frequencies. The color chosen for the histogram is consistent with the color used on GitHub's Insights page for showing commit frequencies.

\section{Case Studies}

To validate the effectiveness of our designed visualizations, we conducted case studies on three public repositories sourced from GitHub: \textit{freeCodeCamp} by \texttt{freeCodeCamp.org}~\cite{freeCodeCamp}, \textit{Hyprland} by Hypr Development~\cite{Hyperland}, and \textit{javascript} by \texttt{airbnb.org}~\cite{airbnb}. The \textit{freeCodeCamp} repository belongs to an online community that supports individuals in learning to code, while the \textit{Hyprland} repository is associated with a dynamic tiling Wayland compositor. On the other hand, the \textit{javascript} repository is maintained by \texttt{airbnb.org} and serves as a JavaScript styling guide. These repositories exhibited variations in terms of programming language, number of contributors and users, and repository management practices. Thus, the case studies served to underscore the efficacy of the proposed visualization solutions across diverse scenarios.

\subsection{freeCodeCamp}
Once the repository data was loaded in the proposed interface, the first page of the application (see \autoref{timeline-freecodecamp}) showed the issues of the repository in the \textsc{Timeline View}~(T0). It could be noticed that the longest open issue, in purple, \texttt{search bar at top of page unable to find challenge} was open for more than 17 days. The longest closed issue, in green, \texttt{CodeAlly Down flag is too wide} took 5 days to be resolved~(T0).

After switching the \textsc{Timeline View} to show labels of the issues (see \autoref{label-freecodecamp}), it could be noticed that the majority of the issues have purple color in it indicating most of the issues were either bug reports or feature requests~(T1). Upon further inspection of the tooltips, it was determined that there were 7 bug-type issues and 8 feature requests. The \texttt{CodeAlly Down flag is too wide} issue that took the longest to get resolved was a bug-type issue. The label also indicated that it required help to solve, which could be a reason why it took longer to fix.

Upon clicking the bars of the \textsc{Timeline View}, the \textsc{Issue Graph} appears. When the bar for \texttt{CodeAlly Down flag is too wide} issue was clicked, the graph view shown in \autoref{graph-freecodecamp-simple} appeared. It can be seen that only two commits were made during the time period of the clicked issue, and both of them contain ``fix" keyword in the message~(T2). Further inspection of commit messages revealed that \texttt{fix: relocate CodeAlly banner to fit layout (\#50534)} commit potentially fixed the issue. We verified that this commit fixed the issue by manually checking it on the GitHub page. The \textsc{Issue Graph} also shows that the commit modified \texttt{client/src/templates/Challenges/codeally/\newline show.tsx} file and was created by \texttt{CallmeHongmaybe}.

\begin{figure}[!thb]
    \centering
    \includegraphics[trim=6pt 5pt 6pt 20pt, clip, width=0.975\linewidth]{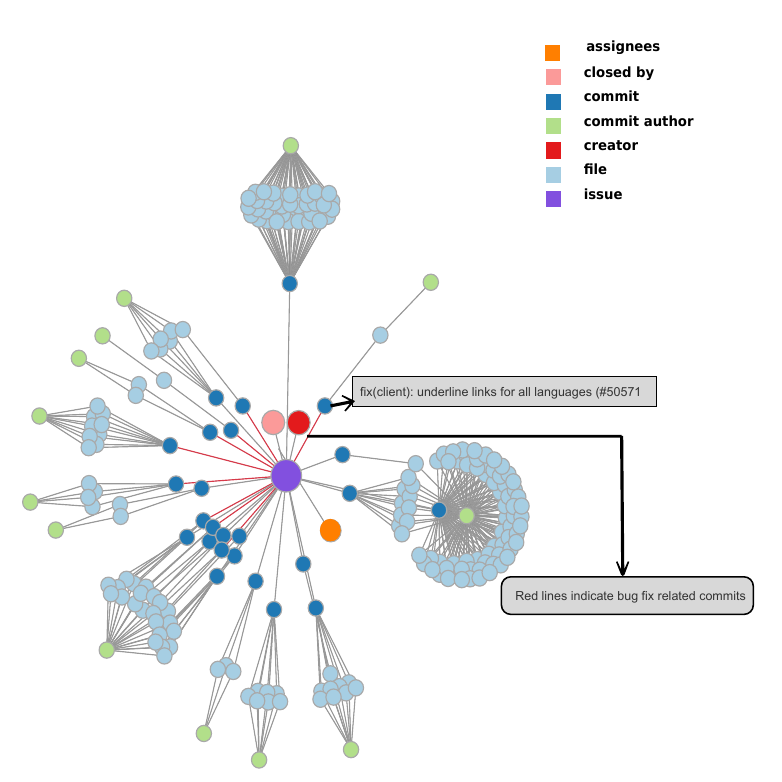}
    \caption{\textsc{Issue Graph} of an open issue titled--\texttt{search bar at top of page unable to find challenge} from \textit{freeCodeCamp} repository. The graph includes the ``closed by" and ``assignee" nodes for demonstration purposes. It should be noted that the issue did not have an assignee, and since it was open, the ``closed by" information is not applicable. All other elements of the graph reflect real data.}
    \label{graph-freecodecamp-complex}
\end{figure}

\autoref{graph-freecodecamp-complex} shows the \textsc{Issue Graph} for the issue that was open for the longest time. The graph pointed out that 25~commits were made since the day the issue was opened. After checking the commit message of all of these commits, it seemed like none of these commits were related to the issue. In the \textsc{Timeline View} showing labels as shown in \autoref{label-freecodecamp}, it showed that the issue was a feature request which is still being discussed. This gives a clearer idea of why the issue was open for a long time. Another interesting aspect of \textsc{Issue Graph} was that it seemed like there were multiple active developers in the repository involved in multiple commits. It was also easier to identify which commits made modifications to multiple files from this graph.

Upon selecting the button to transition to the \textsc{Summary of Updated Files} view, it becomes evident that the file with the highest count of updated lines is the \texttt{pnpm-lock.yaml}  file~(T3), as depicted in \autoref{donut-freecodecamp}(left). This observation aligns with expectations, considering that the freeCodeCamp project remains active and frequently receives updates to its packages, resulting in newer versions being incorporated. The histogram presented in \autoref{donut-freecodecamp}(middle) reveals that most files experienced minor changes, with 2-3 lines of code being updated. This pattern suggests that most files within the repository underwent relatively small modifications.

\subsection{Hyprland}

For the \textit{Hyprland} repository, the temporal aspects of issues are visually represented in \autoref{timeline-hyprland}. This view reveals a noticeable presence of recently opened issues within the repository. Notably, the issue titled \texttt{1 pixel gaps on no-gap settings} remained open for the longest duration, spanning more than 28 days~(T0). Upon switching the \textsc{Timeline View} to display labels, as demonstrated in \autoref{label-hyperland}, it became apparent that all the issues in the repository were assigned to a single label, distinguishing it from the \textit{freeCodeCamp} repository. It is noteworthy that this repository predominantly utilizes default GitHub labels. Moreover, a significant proportion of the issues pertained to bug reports~(T1).

The bug report titled \texttt{Resize window after ungrouping a single window group} stood out as the most time-consuming issue to be resolved. Upon selecting the corresponding bar in the \textsc{Timeline View}, it's \textsc{Issue Graph} was displayed, as illustrated in \autoref{graph-hyprland}. The graph provides an overview of the issue lifecycle, revealing a total of 25 commits that were made during the lifecycle. Notably, the user \texttt{vaxerski} was attributed to 21 of these commits. Upon examining the commit messages, one specific commit with the message \texttt{window: recalc on deco remove} emerged as a potential fix for the issue~(T2). To validate its significance, verification was performed using GitHub's user interface, confirming its role in addressing the issue. The graph further highlighted that the commit exclusively modified the \texttt{src/Window.cpp} file and was authored by the user \texttt{vaxerski}. Particularly, the graph emphasized the substantial contribution made by the user \texttt{vaxerski}, suggesting their significant involvement as one of the main contributors to the repository.

\begin{figure*}[!tbp]
    \centering

    \subfigure[\textsc{Timeline View} of issue status\label{timeline-javascript}]{{\includegraphics[trim=100pt 20pt 60pt 5pt, clip, height=105pt]{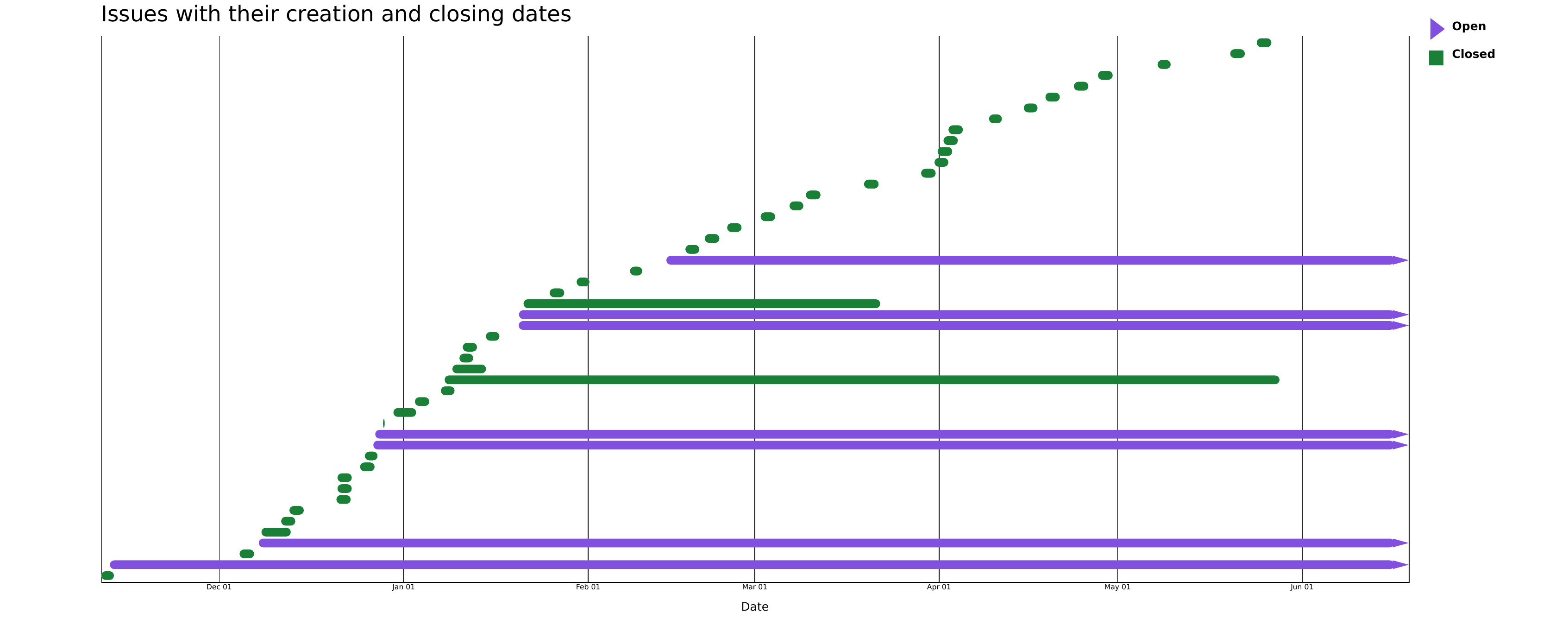}}}
    \hfill
    \subfigure[\textsc{Timeline View} of issue labels\label{label-javascript}]{{\includegraphics[trim=30pt 20pt 10pt 10pt, clip, height=105pt]{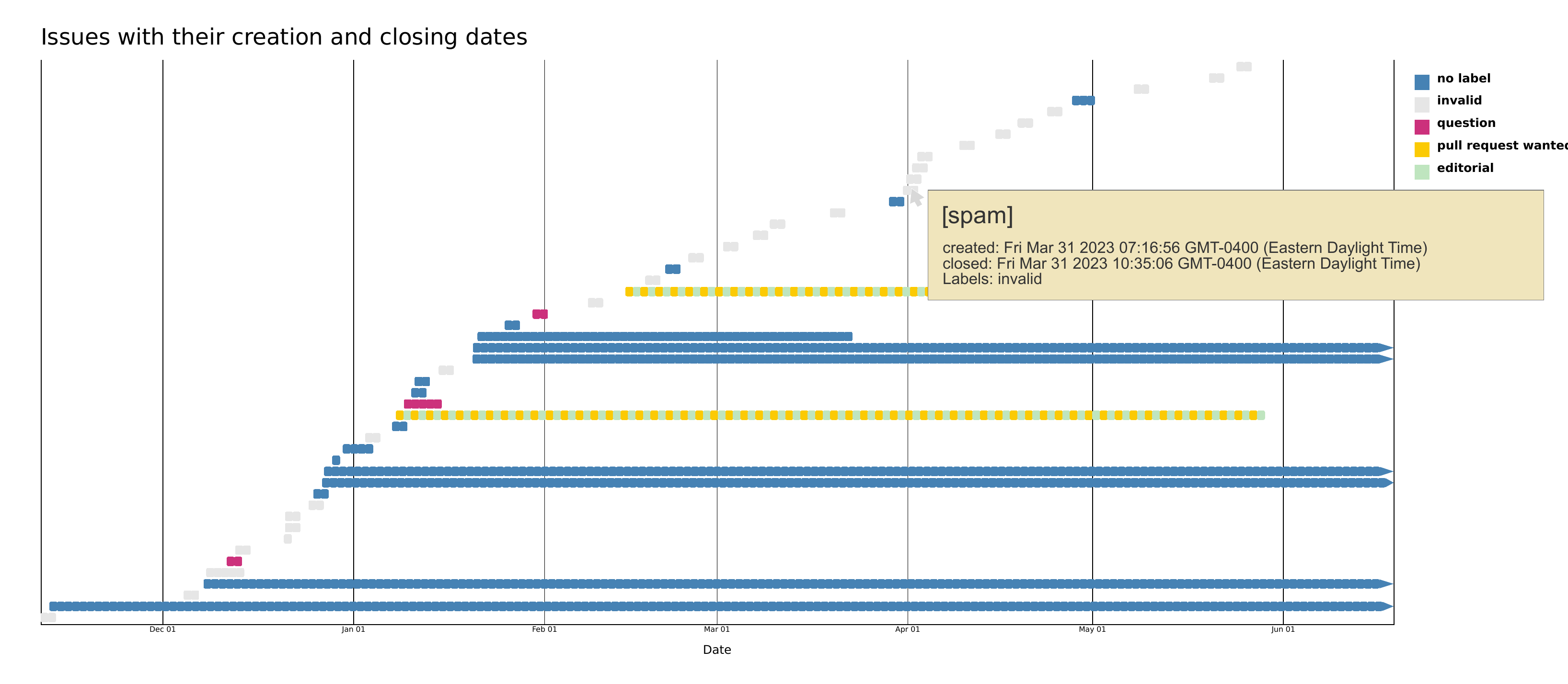}}}

    \subfigure[Top five most updated files\label{donut-javascript}]{{\includegraphics[trim=5pt 40pt 10pt 35pt, clip, width=0.25\linewidth]{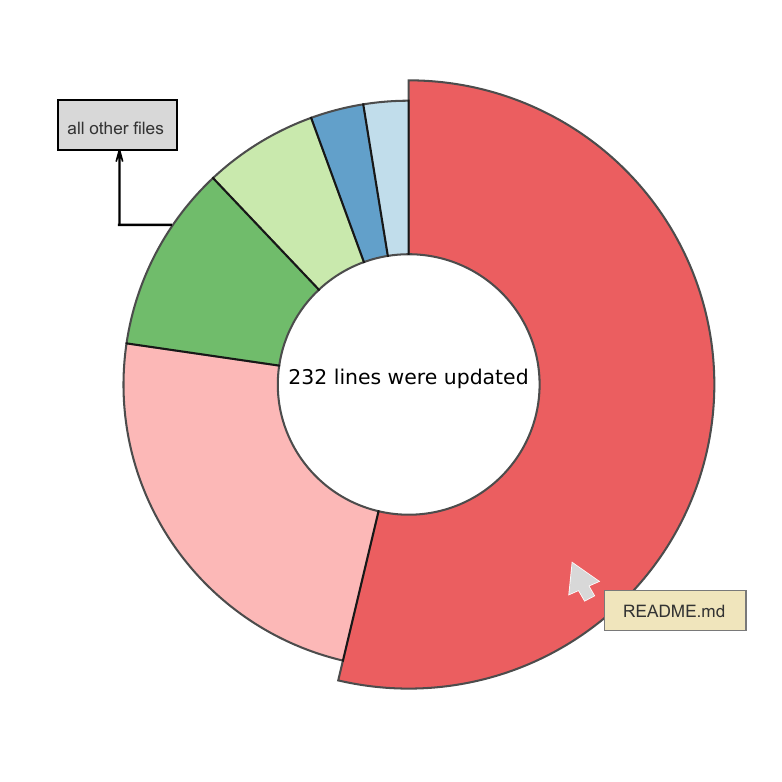}}}
    \hspace{5pt}
    \subfigure[Distribution of number of files across different levels of modification\label{histogram-javascript}]{{\includegraphics[trim=20pt 0pt 40pt 5pt, clip, width=0.7\linewidth]{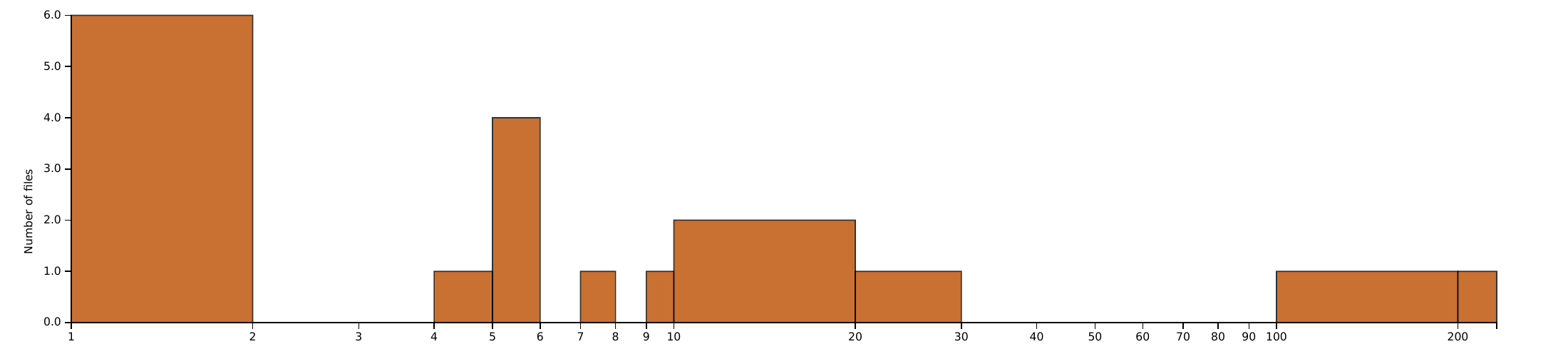}}}

    \caption{Visualizations of the issues created between November 11, 2022, and June 18, 2023, in \textit{JavaScript} repository. The \textsc{Timeline View} in (a) illustrates that many issues were closed quickly. The \textsc{Timeline View} in (b) indicates that almost all of them are labeled invalid. The donut chart in (c) displays the distribution of updated lines among different files. Notably, the \texttt{README.md} file stands out with significantly more modified lines than other files in the repository. The histogram in (d) indicates that most files experienced only minor updates, typically involving 1-2 lines.}
    \label{javascript-all}
\end{figure*}

The \textsc{Summary of Updated Files} view for the \textit{Hyprland} repository is showcased in \autoref{donut-hyprland}. Among the files, the \texttt{src/debug/HyprCtl.cpp} file experienced the highest activity level, with 23 lines of code being updated, making it the most updated file within the repository~(T3). Additionally, when considering all the remaining files collectively, a total of 30 lines underwent modifications. By applying the filter to unveil files updated during bug-fixing commits, the resulting donut chart illustrated in \autoref{donut-bug-hyprland} revealed that the file named \texttt{src/Windows.cpp} underwent updates on 19 lines during bug-fix related commits.

\subsection{Javascript}

The \textsc{Timeline View} of the \textit{javascript} repository is presented in \autoref{timeline-javascript}. From this view, it is evident that several recent issues were resolved within a short timeframe~(T0). Upon switching to the label view, it was observed that a significant number of these issues were labeled as \texttt{invalid}~(T1). Further investigation through tooltips revealed that these issues were flagged as spam. Out of the total 50 issues, 28 were marked as spam, indicating a recent influx of spam-related challenges in the repository. The label view in \autoref{label-javascript} also highlights that none of these issues were bug reports, and a considerable portion of them was left unassigned.

The \textsc{Summary of Updated Files} view, depicted in \autoref{donut-javascript}, highlights the file \texttt{README.md} as the most frequently modified file, with 232 lines updated. The green section in the figure represents the lines updated in all other files, excluding the top five most updated files, which accounted for 46 lines. The notable disparity between \texttt{README.md} and the remaining files suggests that the repository primarily focused on documentation updates~(T3). The accompanying histogram in \autoref{histogram-javascript} indicates that most files in the repository underwent minor changes, with 1-2 lines updated. However, two files experienced a more substantial number of modifications, with a range of 100-232 lines updated.

\begin{figure}[!t]
    \centering

    \includegraphics[width =0.5\textwidth]{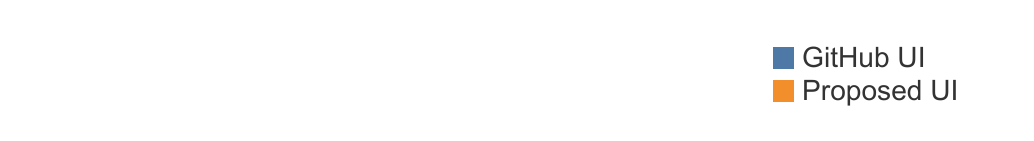}
    
     \subfigure[Time taken to answer Q1\label{Q1-result}]                    {\includegraphics[width=0.415\linewidth]{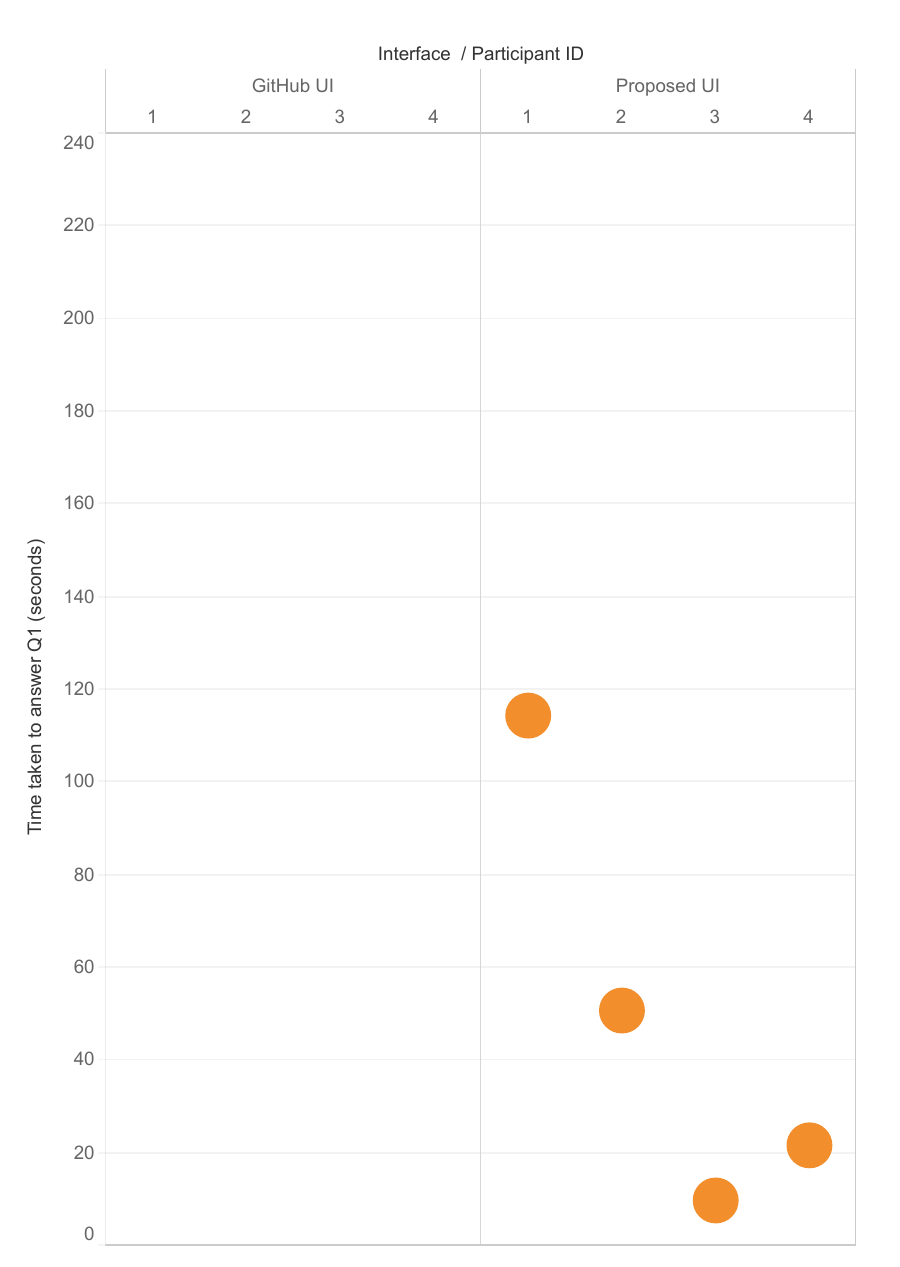}}
    \hspace{5pt}
     \subfigure[Time taken to answer Q2\label{Q2-result}]        {\includegraphics[width=0.415\linewidth]{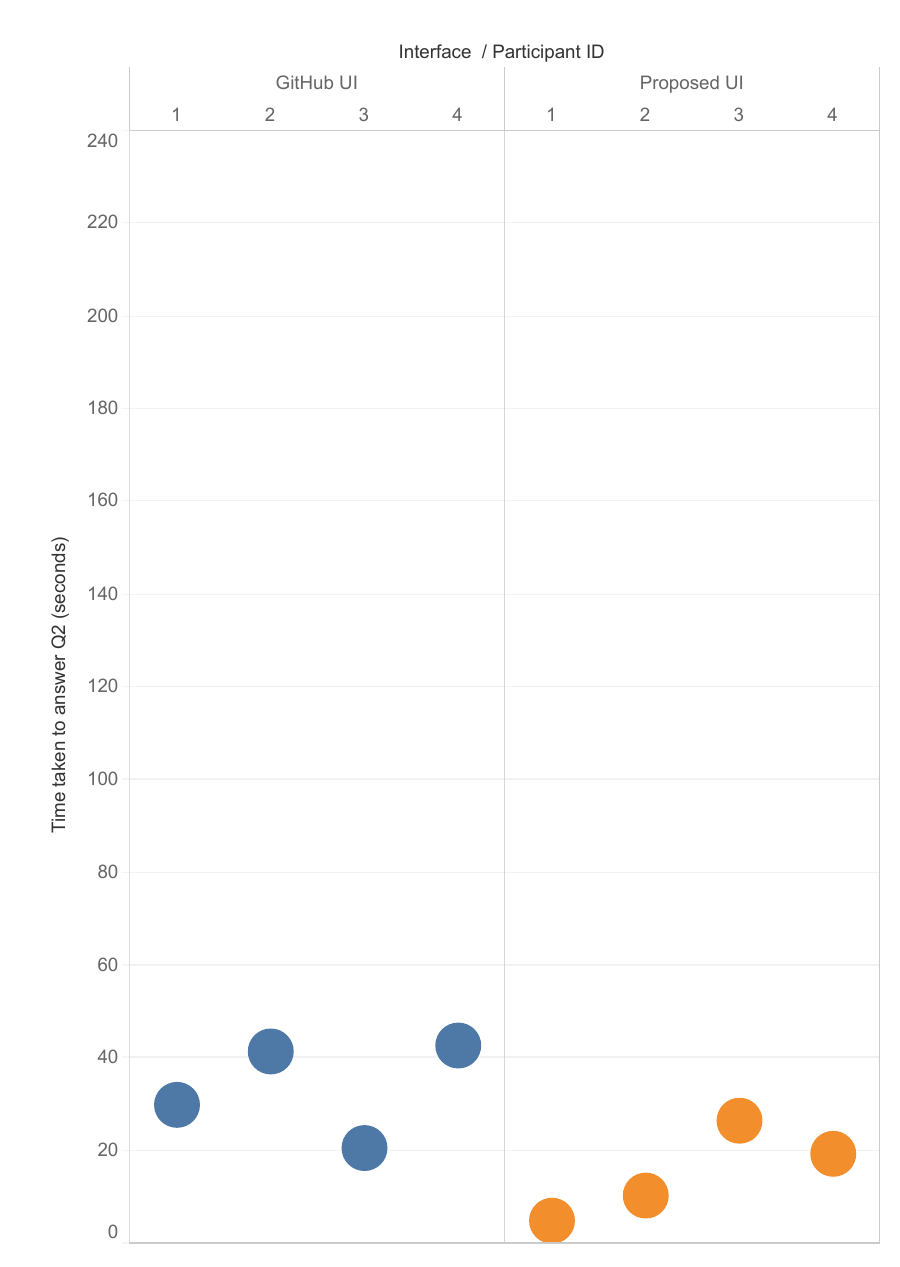}}

    \subfigure[Time taken to answer Q3\label{Q3-result}]                    {\includegraphics[width=0.415\linewidth]{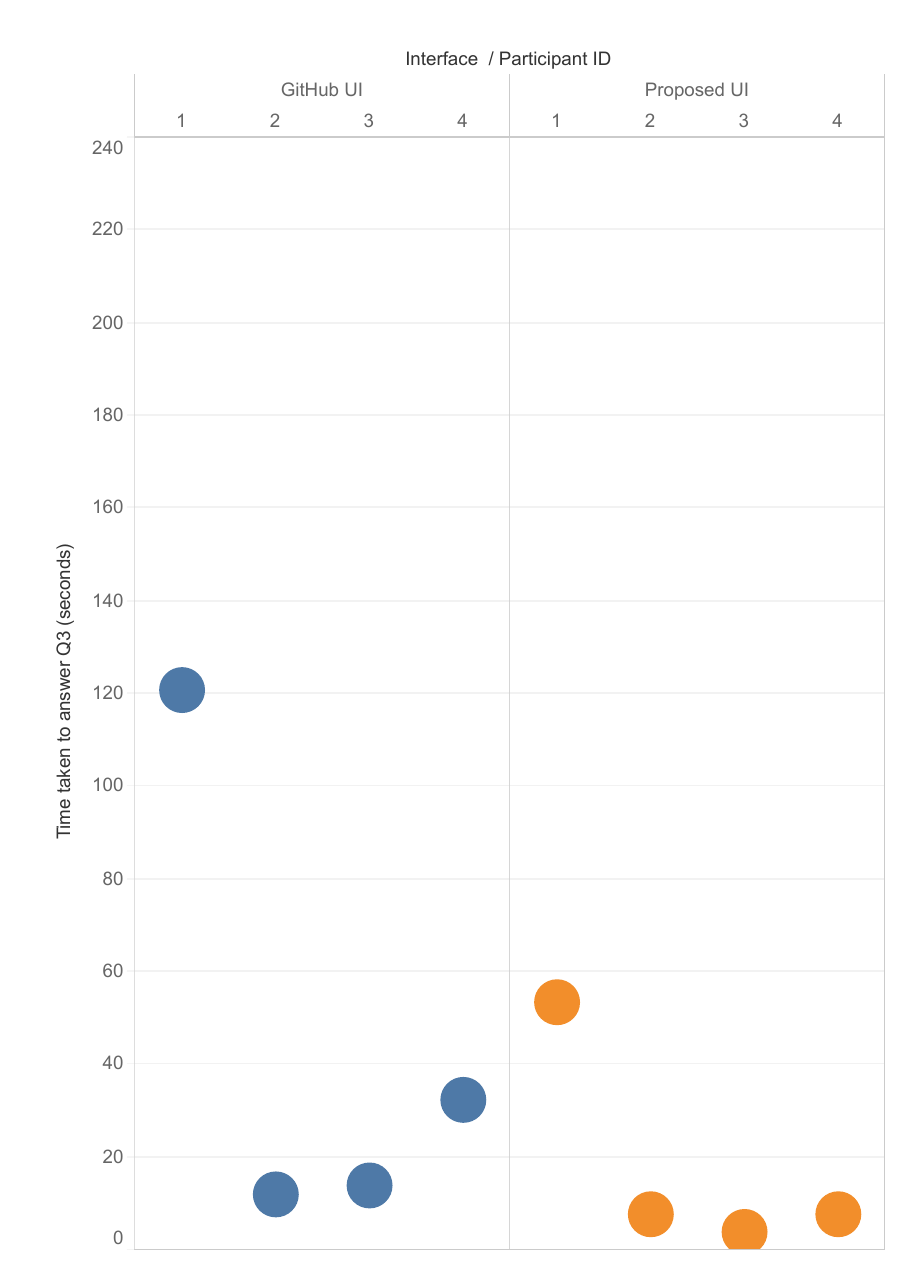}}
    \hspace{5pt}
     \subfigure[Time taken to answer Q4\label{Q4-result}]        {\includegraphics[width=0.415\linewidth]{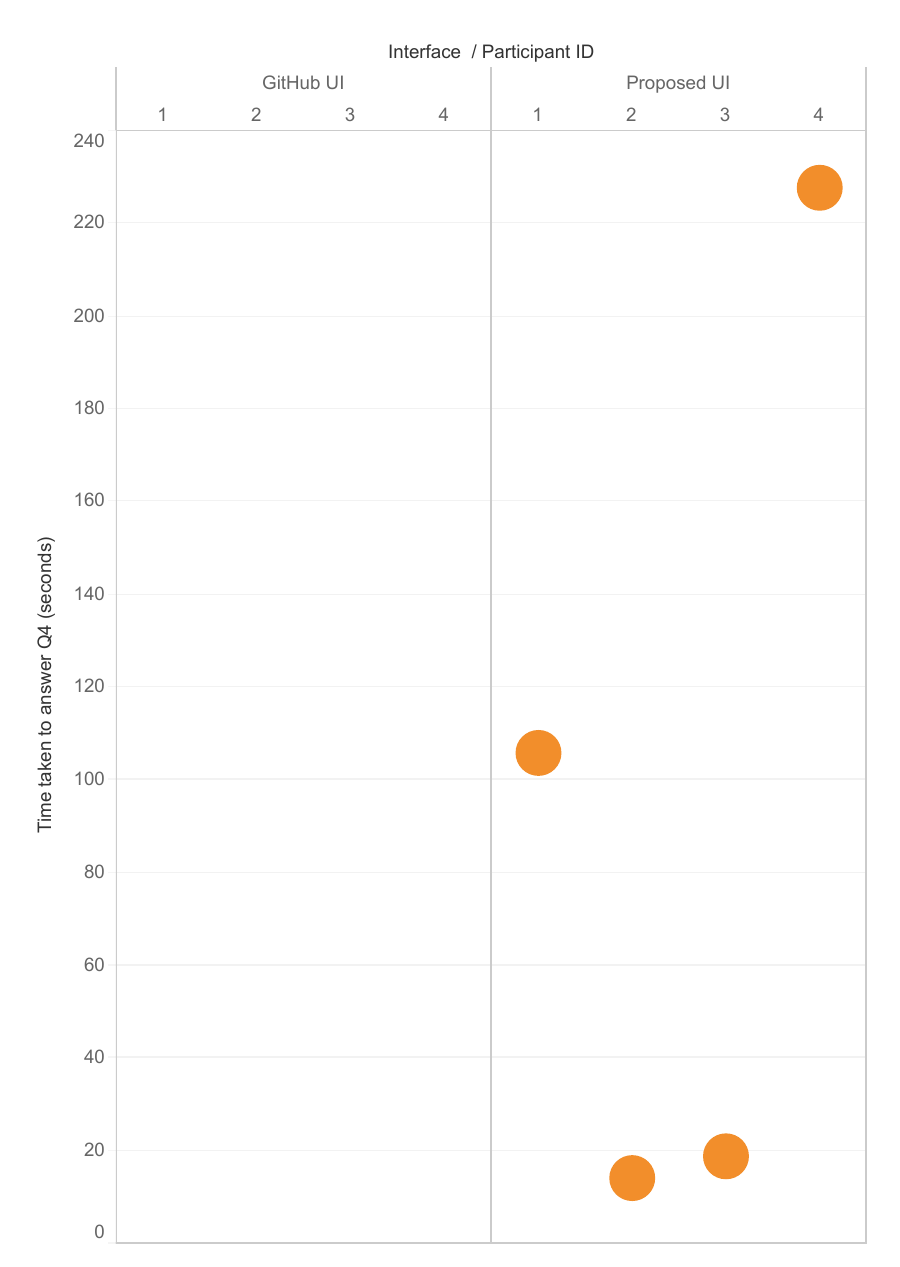}}

    \subfigure[Time taken to answer Q5\label{Q5-result}]                    {\includegraphics[width=0.415\linewidth]{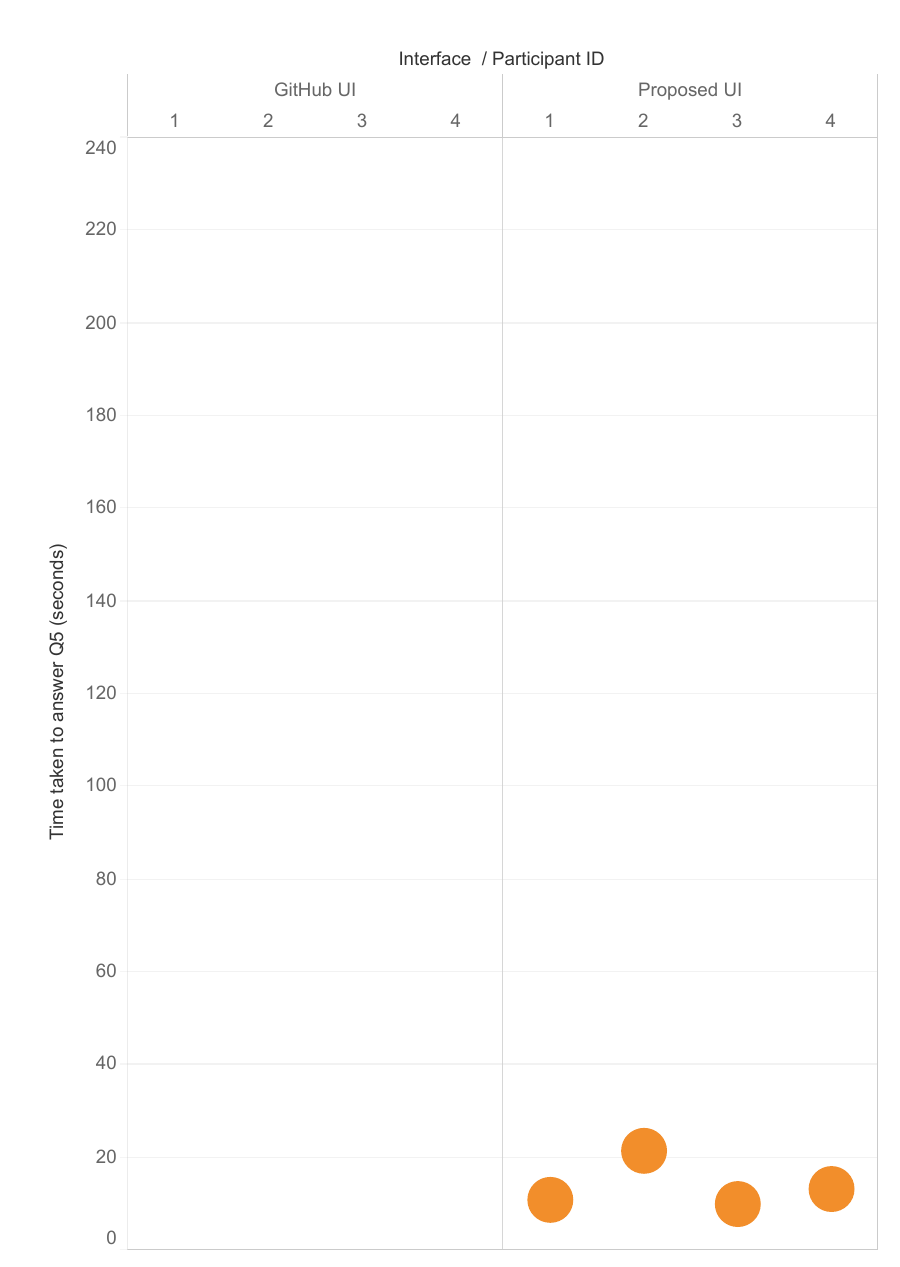}}
    \hspace{5pt}
     \subfigure[Time taken to answer Q6\label{Q6-result}]        {\includegraphics[width=0.415\linewidth]{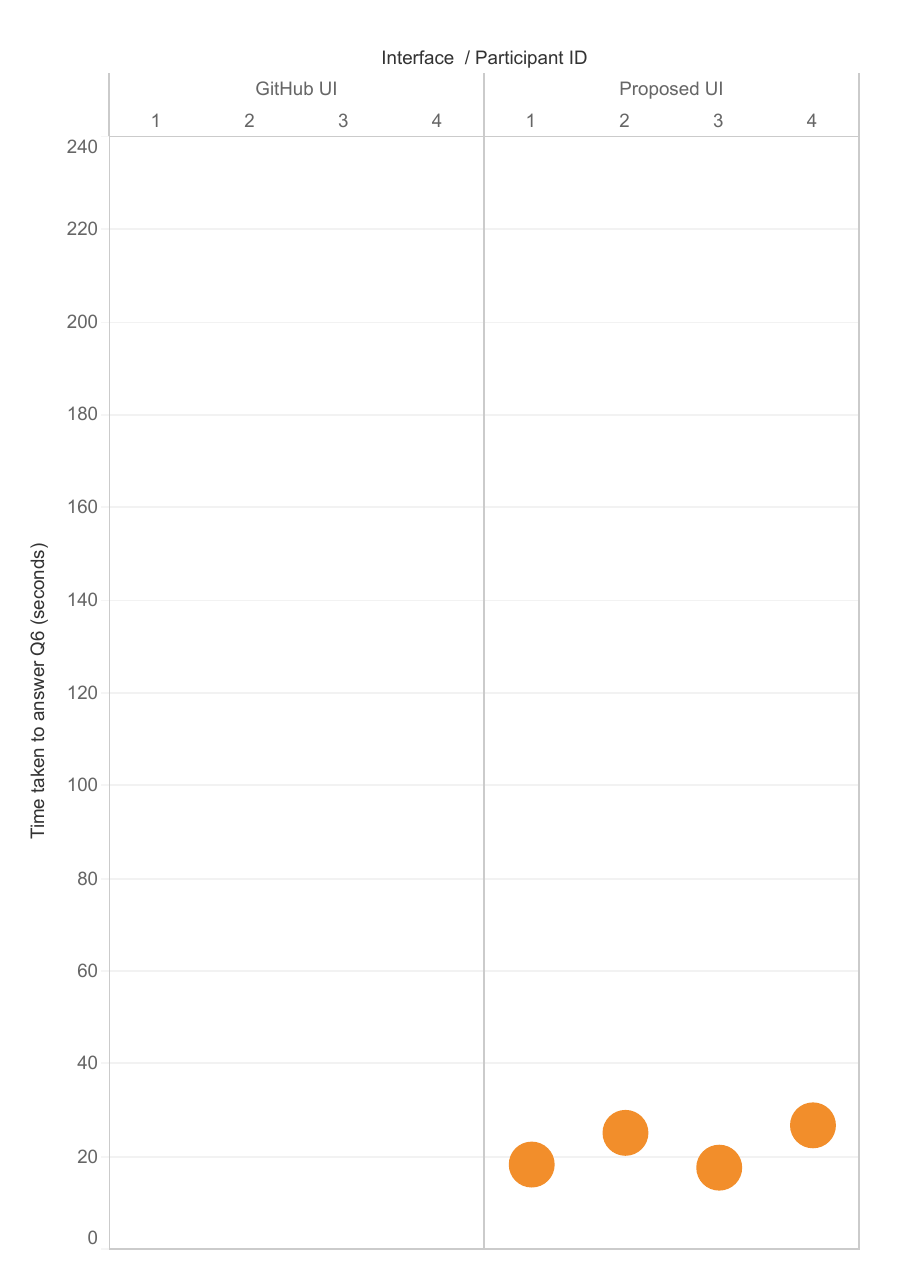}}

    \caption{The user evaluation results compare the proposed interface and the GitHub user interface. Notably, certain charts lack GitHub markings due to participants reporting unable to answer those questions without investing considerable time. Further, for \autoref{Q2-result} and \autoref{Q3-result}, participants took less time on average to answer using the proposed interface than the GitHub user interface.}
    \label{results}
\end{figure}

\section{User Evaluation}
To evaluate the effectiveness of the proposed visualizations achieving the tasks described in~\ref{sec:design} compared to the existing GitHub interface, we conducted a small-scale user evaluation involving 4 participants, including two professional software developers and two graduate students majoring in computer science and engineering. Among the software developers, one had prior involvement in the development process, while the other had no prior exposure to the tool before the evaluation. All participants had prior experience using GitHub. In terms of measuring effectiveness, we mainly focused on the time participants took to answer certain questions using the proposed visualizations and the existing GitHub UI.

\subsection{Evaluation Design}
We conducted online interviews with the participants, giving them a brief overview of the study's purpose. They were also given a demonstration on how to use the proposed web tool on a repository different from the one on which they were assigned tasks. Additionally, we showcased essential features of the actual GitHub user interface (UI), such as sorting issues based on different parameters, checking labels, accessing information about specific issues and commit pages of a repository, and retrieving details regarding file updates. Once the participant felt comfortable using our proposed web tool and the GitHub UI, we proceeded with the evaluation.

For the evaluation, participants were tasked with answering a set of questions for a given repository using the proposed interface and answering the same questions for a different repository using the GitHub UI. This approach was designed to minimize bias and facilitate the comparison of learning patterns. To assess performance, we measured participants' time to answer each question using a timer for both the proposed interface and the GitHub UI. The timer was started once the question was asked and stopped once participants found an answer.

For the evaluation process, we chose two repositories, namely \textit{freeCodeCamp} and \textit{Hyprland}. Two participants interacted with the \textit{freeCodeCamp} repository using our proposed interface, while they used the GitHub UI for the \textit{Hyprland} repository. Conversely, the other two participants swapped repositories, using the proposed interface for \textit{Hyprland} and the GitHub UI for \textit{freeCodeCamp}. Participants were asked a series of questions designed to measure the effectiveness of the proposed interface to accomplish the tasks defined in~\ref{sec:design}. The questions are:

    \begin{enumerate}[start=1,label={Q\arabic*:}]
        \item Which closed issue took the longest time to get resolved? Who opened it? Who closed it?
        \item Which issue has been open for the longest time?
        \item Which labeled issue is the majority in this repository?
        \item Which bug-related issue took the longest to get resolved? Who opened it? Who closed it? Which commit resolved it?
        \item Which file has been updated the most?
        \item Which file has been updated the most for bug fixes?
    \end{enumerate}

\subsection{Results}

 The time taken by the participants to answer the questions is shown in \autoref{results}. The missing values for GitHub in \autoref{Q1-result}, \autoref{Q4-result}, \autoref{Q5-result}, and \autoref{Q6-result} denote that the participant reported that answering the question using the interface will take a significant amount of time, and thus they could not provide an answer. However, the questions that were answered were correct. For Q1 and Q4, the participants reported that the GitHub UI does not display the duration it took to close an issue. Therefore, they would need to calculate each issue manually. Additionally, GitHub utilizes terms like ``last week," ``two weeks ago," etc., to indicate when a specific issue was opened or closed, making it challenging to determine the exact duration. For Q5 and Q6, participants had to manually inspect each file in the repository to determine the number of updated lines, requiring meticulous file tracking. This cumbersome process makes it difficult to provide answers within a reasonable time frame. However, for Q2 and Q3, the participants, on average, required less time when using the proposed interface than the GitHub UI.

\section{Conclusion \& Future Work}

This paper introduces a set of visualizations designed to facilitate the analysis of repositories by programmers, specifically focusing on issues and their corresponding commits. The objective is to provide programmers with a straightforward means of deriving meaningful insights to enhance project quality.

Through our case study and user evaluation, we have observed that these visualizations effectively present information, enabling users to gain valuable insights quickly. One notable feature of our approach is the ability to identify files that have been updated to address bug fixes. This functionality can be expanded to encompass updates on all labels within a repository. Furthermore, while our current system detects updates at the file level, incorporating more advanced logic to identify updates at the function or code level would allow for a deeper understanding of the relationships between these elements and issues. 

In our future research, we intend to analyze comments associated with issues to extract additional valuable insights. These insights will be presented through visualizations, providing a more efficient and expedient means of exploration. It is worth noting that GitHub issues have the potential to be reopened, and our current approach does not account for this scenario in identifying or detecting reopened issues. Furthermore, we acknowledge that the current interface lacks the ability for users to view issues within a specific time interval. Incorporating these functionalities and validating them with a more extensive larger scale user evaluation is an important aspect of our future plans.

Involving professional software developers in the design and evaluation process demonstrates the usability of our proposed approach in the software development field. Incorporating their input allowed us to refine and tailor the design to meet their specific needs. We strongly believe that this tool holds great potential to assist the programming community in gaining a comprehensive understanding of their projects and identifying areas for improvement. 

High-resolution versions of all the figures in this paper can be accessed at: \texttt{\small\url{https://osf.io/bx8hs/?view_only=4cab290284e542708c4c4b100de12de2}}.

\bibliographystyle{IEEEtran}
\bibliography{ref}

\end{document}